\documentclass[12pt]{article}

\usepackage{cite}
\usepackage{epsfig}
\usepackage{color}

\input paperdef

\begin{document}
\thispagestyle{empty}

\def\thefootnote{\fnsymbol{footnote}}

\begin{flushright}
CERN-TH/2003-91\\
DCPT/03/62\\
IPPP/03/31\\
LMU 06/03\\
ZU-TH 05/03\\
hep-ph/0306181\\
\end{flushright}

\vspace{.2cm}

\begin{center}

{\large\sc {\bf Physics Impact of a Precise Determination}}

\vspace{0.4cm}

{\large\sc {\bf of the Top Quark Mass at an} $\mathbf{e^+e^-}$ 
                {\bf Linear Collider}}
 
\vspace{1cm}

{\sc 
S.~Heinemeyer$^{1}$%
\footnote{email: Sven.Heinemeyer@physik.uni-muenchen.de}%
, S.~Kraml$^{2}$%
\footnote{email: Sabine.Kraml@cern.ch}%
, W.~Porod$^{3}$%
\footnote{email: porod@physik.unizh.ch}%
~and G.~Weiglein$^{4}$%
\footnote{email: Georg.Weiglein@durham.ac.uk}
}

\vspace*{1cm}

{\sl
$^1$Institut f\"ur theoretische Elementarteilchenphysik,
LMU M\"unchen, Theresienstr.\ 37, D-80333 M\"unchen, Germany

\vspace*{0.4cm}

$^2$CERN, TH Division, CH-1211 Geneva 23, Switzerland

\vspace*{0.4cm}

$^3$Institut f\"ur Theoretische Physik, Universit\"at Z\"urich,\\
CH-8057 Z\"urich, Switzerland

\vspace*{0.4cm}

$^4$Institute for Particle Physics Phenomenology, University of Durham,\\
Durham DH1~3LE, UK

}

\end{center}

\vspace*{1cm}

\begin{abstract}
At a prospective \epem\ Linear Collider (LC) a very precise 
determination of the top quark mass with an accuracy of
$\de\mt \lsim 100 \mev$ will be possible.
This is to be compared with the envisaged accuracy of 
$\de\mt = 1$--$2 \gev$ at the Tevatron and the LHC. 
We discuss the physics impact of such a precise determination of $\mt$, 
focusing on the Standard Model (SM) and its minimal supersymmetric 
extension (MSSM). 
In particular, we show the importance of a precise knowledge of $\mt$ 
for electroweak precision observables, and for Higgs physics and the
scalar top
sector of the MSSM. Taking the mSUGRA model
as a specific example, we furthermore demonstrate the importance of a
precise $\mt$ value for the prediction of sparticle masses and for
constraints on the parameter space allowed by the relic density. The
uncertainty in $\mt$ also significantly affects the reconstruction of the 
supersymmetric high scale theory. We find that going from hadron
collider to 
LC accuracy in $\mt$ leads to an improvement of the investigated
quantities by up to an order of magnitude.
\end{abstract}

\def\thefootnote{\arabic{footnote}}
\setcounter{page}{0}
\setcounter{footnote}{0}

\newpage


\section{Introduction}

At a prospective \epem\ Linear Collider (LC) a very precise determination
of the top quark mass with an accuracy of
\BE
\de\mtexp \; \lsim \; 100 \mev {\rm ~~(LC)}
\label{eq:mtlc}
\EE
will be possible~\cite{tesla,nlc,jlc,mtdet1,mtdet2,mtdet3}. 
This has to be compared with the envisaged accuracy of  
\BE
\de\mtexp \; = \; \mbox{1--2} \gev {\rm ~~(Tevatron, LHC)}
\EE
at hadron colliders, i.e.\ Run~II of the Tevatron and the LHC~\cite{mtdetLHC}. 
The question arises of what can be learnt from the improved accuracy
obtainable at the LC. Some implications have recently been discussed
in \citere{Chakraborty:2003iw}. 
In the present paper, we perform a detailed investigation of the physics impact 
of the LC precision on $\mt$, focusing on the Standard 
Model (SM) and the Minimal Supersymmetric Standard Model (MSSM). 
We study the dependence of different quantities on $\mt$ and compare 
these effects with the anticipated future experimental accuracies, 
taking into account theoretical uncertainties induced by the
experimental errors of other input parameters, as well as uncertainties
from unknown higher-order corrections.

We discuss in particular possible improvements in the
analysis of electroweak precision observables induced by a precision
measurement of $\mt$. Moreover, we demonstrate the impact of the
achievable precision in $\mt$ on the phenomenology of the Higgs and
stop sectors of the MSSM, on the possibility to reconstruct the
underlying high scale theory, and on constraints from the dark matter
relic density.

While the examples presented in this paper are by no means exclusive,
we believe that they give 
a fair idea 
of the physics impact of a very precise determination of $\mt$. 
Other examples where $\mt$ is an important parameter are, for example,  
$B$~and $K$~physics. $B$~and $K$~decays and $B^0$--$\bar B^0$ mixing are, 
however, significantly affected by hadronic uncertainties, such that a 
$\mt$ precision of a few GeV is sufficient; see also 
\citere{Chakraborty:2003iw}. 
Also for the high-precision determination of $\als(\MZ)$ at 
the $Z$-boson resonance (GigaZ) it turned out 
that $\de \mt\sim 1 \gev$ will be sufficient~\cite{alsdet}. 

In our analysis we do not discuss different definitions of the top quark mass.
The accuracy of the mass parameter extracted from the $t \bar t$ threshold 
measurements at the LC will be significantly better than the $100 \mev$ 
quoted in
\refeq{eq:mtlc}, see \cite{tesla,nlc,jlc,mtdet1,mtdet2,mtdet3}.
However, its transition into a short-distance mass (like the \msbar\ mass) 
that is suitable as an input parameter for the observables 
that we investigate below, involves 
further theoretical uncertainties. Taking these uncertainties into
account, an accuracy in the $\mt$ determination of $\de\mt \lsim 100
\mev$ as quoted in \refeq{eq:mtlc} seems to be feasible (for a review,
see \cite{mtdet1} and references therein). This requires an
experimental accuracy of $\als(\MZ)$ of about 0.001, which can be
obtained from LC measurements also at the $t \bar t$ threshold~\cite{mtdet2}, 
possibly from event shape measurements at the LC~\cite{mtdet3}, and 
from LC measurements at GigaZ, where an even
higher accuracy seems to be achievable~\cite{alsdet}.

The remaining parts of the paper are organized as follows:
\Refse{sec:EWPO} focuses on the electroweak precision observables in
the SM and the MSSM.
Internal consistency checks of the two models are investigated, and the 
impact of the precision in $\mt$ is analysed with respect to the
experimental accuracy of the precision observables and the theoretical
uncertainties of the predictions. 
The influence of $\de\mt$ on precision physics in the MSSM 
Higgs and stop sectors is analysed in
\refse{sec:MSSMimp}. In \refse{sec:RG} we study the relevance of
$\de \mt$ in renormalization group (RG) running of SUSY parameters.
We investigate the consequences for chargino and neutralino mass
predictions, dark matter constraints on mSUGRA parameters, and for the
reconstruction of the SUSY--breaking boundary conditions.
\Refse{sec:conclusions} gives a summary and conclusions.


\section{Electroweak precision observables}
\label{sec:EWPO}

\subsection{Consistency tests of the SM and the MSSM}
\label{subsec:consistency}

Electroweak precision observables (EWPO) can be used to
perform internal consistency checks of the model under consideration and
to obtain indirect constraints on the unknown model parameters.
This is done by comparing the experimental results of the EWPO with their
theory prediction within, for example, the SM or the MSSM.
In this subsection, we focus 
on the two most prominent observables in the context of electroweak
precision tests, the $W$~boson mass $\MW$ and
the effective leptonic mixing angle $\sweff$. 

Currently the uncertainty in $\mt$ is by far the dominant effect in
the theoretical uncertainties of the EWPO. Today's experimental errors
of $\MW$ and $\sweff$~\cite{ewdataw03} are shown in
\Refta{tab:ewpounc}, together with the prospective future experimental
errors 
obtainable at the Tevatron~\cite{ewpo:tev}, the LHC~\cite{ewpo:lhc},
and the LC without~\cite{tesla,nlc,ewpo:lc} and with a GigaZ
option~\cite{ewpo:gigaz1,ewpo:gigaz2} (see \cite{blueband} for a
compilation of these errors and additional references).

\begin{table}[htb!]
\renewcommand{\arraystretch}{1.5}
\begin{center}
\begin{tabular}{|c||c|c|c|c|}
\cline{2-5} \multicolumn{1}{c||}{}
& Today & Tevatron/LHC & ~LC~  & GigaZ \\
\hline\hline
$\de\sweff(\times 10^5)$ & 17 & 14--20   & --  & 1.3  \\
\hline
$\de\MW$ [MeV]           & 34 & 15   & 10   & 7      \\
\hline\hline
\end{tabular}
\end{center}
\caption{Experimental errors of $\MW$ and $\sweff$ at present and future
  colliders~\cite{ewdataw03,ewpo:tev,ewpo:lhc,tesla,nlc,ewpo:lc,ewpo:gigaz1,ewpo:gigaz2,blueband}. 
}
\label{tab:ewpounc}
\renewcommand{\arraystretch}{1}
\end{table}

There are
two sources of theoretical uncertainties: (i) 
those from
unknown higher-order corrections, which we call intrinsic theoretical
uncertainties, and (ii) 
those from experimental errors of the input
parameters, which we call parametric theoretical uncertainties. The
intrinsic uncertainties within the SM are currently~\cite{blueband,georgtalk}
\BE
\De\MW^{\rm intr,today} \approx \pm 4 \mev, \quad
\De\sweff^{\rm intr,today} \approx \pm 6 \times 10^{-5}~.
\label{eq:intruncSM}
\EE
They are based on the present status of the theoretical predictions in
the SM, namely the complete two-loop result for $\MW$ (see
\cite{MWtwoloop} and references therein), a
partial two-loop result for $\sweff$ (see \cite{dgs} and references
therein) and leading three-loop contributions to both observables (see
\cite{faisst} for the latest result, and references therein). For the
MSSM, the available results beyond one-loop order are less advanced than
in the SM (for the latest two-loop results, see \cite{ytdet} 
and references therein). Thus, the intrinsic uncertainties in the MSSM
are still considerably larger than the ones quoted for the SM in 
\refeq{eq:intruncSM}~\cite{ewpoSUSYprocs}.

The parametric uncertainties induced by the experimental errors of
$\mt$ and $\De\al_{\rm had}$ are currently~\cite{MWradcor}
\BEA
\de\mt = 5.1 \gev &\Rightarrow& 
\De\MW^{{\rm para},\mt} \approx \pm 31 \mev, \quad
\De\sweff^{{\rm para}, \mt} \approx \pm 16 \times 10^{-5} \\[.3em]
\de(\De\al_{\rm had}) = 36 \times 10^{-5} &\Rightarrow&
\De\MW^{{\rm para},\De\al_{\rm had}} \approx \pm 6.5 \mev, \quad
\De\sweff^{{\rm para},\De\al_{\rm had}} \approx 
  \pm 13 \times 10^{-5}~. \non
\EEA
Accordingly, the parametric uncertainties of $\MW$ and $\sweff$
induced by $\de\mt$ are
approximately as large as the current experimental errors. For the
case of $\MW$ the parametric uncertainty $\De\MW^{{\rm para},\mt}$ is
more than four times larger than 
$\De\MW^{{\rm para},\De\al_{\rm had}}$ and more than 15 times larger 
than $\De\MW^{\rm intr,today}$. 

In order to exploit the high experimental precision 
of the EWPO obtainable at the next generation of colliders
for constraining effects of new physics, a precise measurement of
$\mt$ is mandatory. 
The parametric uncertainties induced by future $\mt$ measurements 
in comparison with the ones
from a future error of $\De\al_{\rm had}$~\cite{fredl} and
from the error of $\MZ$ are~\cite{mtdetLHC,ewpoSUSYprocs}
\BEA
\de\mt = 2 \gev &\Rightarrow& 
\De\MW^{{\rm para},\mt} \approx \pm 12 \mev, \quad
\De\sweff^{{\rm para}, \mt} \approx \pm 6 \times 10^{-5} \non \\[.3em]
\de\mt = 1 \gev &\Rightarrow& 
\De\MW^{{\rm para},\mt} \approx \pm 6 \mev, \quad
\De\sweff^{{\rm para}, \mt} \approx \pm 3 \times 10^{-5} \non \\[.3em]
\de\mt = 0.1 \gev &\Rightarrow& 
\De\MW^{{\rm para},\mt} \approx \pm 1 \mev, \quad
\De\sweff^{{\rm para}, \mt} \approx \pm 0.3 \times 10^{-5}
\label{eq:ewpounc}\\[.3em]
\de(\De\al_{\rm had}) = 5 \times 10^{-5} &\Rightarrow&
\De\MW^{{\rm para},\De\al_{\rm had}} \approx \pm 1 \mev, \quad
\De\sweff^{{\rm para},\De\al_{\rm had}} \approx \pm 1.8 \times 10^{-5}
                                                              \non\\[.3em]
\de\MZ = 2.1 \mev &\Rightarrow& 
\De\MW^{{\rm para},\MZ} \approx \pm 2.5 \mev, \quad
\De\sweff^{{\rm para}, \MZ} \approx \pm 1.4 \times 10^{-5}~. \non 
\EEA
In order to keep the theoretical uncertainty induced by $\mt$ at 
a level comparable to or smaller than the other parametric and intrinsic
uncertainties (the latter are expected to further improve in the
near future), $\de\mt$ has to be smaller than about $0.2 \gev$ in the
case of $\MW$, and about $0.5 \gev$ in the case of $\sweff$.%
\footnote{The parametric theoretical uncertainty of $\sweff$ appears to
be limited by the prospective accuracy of $\De\al_{\rm had}$ in  
\refeq{eq:ewpounc}. However, the theoretical uncertainty of $\sweff$ can 
be further reduced in this case by trading $\De\al_{\rm had}$ as an input
parameter for the Fermi constant $\gf$ and $\MW$, see \refeq{eq:Deltar}
below, leading in this way to an even stricter requirement on the
experimental precision of $\mt$.}
This level of theoretical accuracy is necessary in order to exploit the
prospective experimental precision of the EWPO at a LC with GigaZ
option, see \Refta{tab:ewpounc}.

\begin{figure}[htb!]
\begin{center}
\epsfig{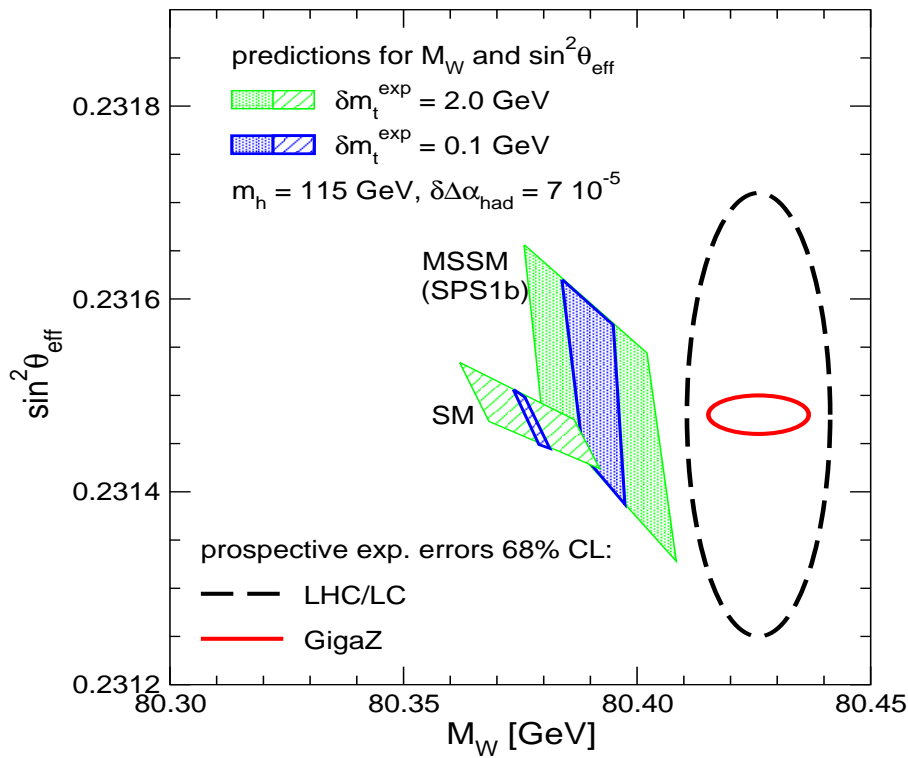}
\caption{
The predictions for $\MW$ and $\sweff$ in the SM and the MSSM
(SPS1b). The inner 
(blue) areas correspond to $\de\mtexp = 0.1 \gev$ (LC), while
the outer (green) areas arise from $\de\mtexp = 2 \gev$ (LHC). 
The anticipated experimental errors on $\MW$ and $\sweff$ at the
LHC/LC and at a LC with GigaZ option are indicated. 
}
\label{fig:SWMW}
\end{center}
\end{figure}

As an example for the potential of a precise measurement of the EWPO
to explore the effects of new physics, we show in \reffi{fig:SWMW} the
predictions for $\MW$ and $\sweff$ in the SM and the MSSM in
comparison with the prospective experimental accuracy obtainable at
the LHC and an LC without GigaZ option (labelled as LHC/LC) and with the 
accuracy obtainable at an LC with GigaZ option (labelled as GigaZ).
The current experimental values are taken as the central ones~\cite{ewdataw03}.
For the Higgs boson mass a future measured value of $\mh = 115 \gev$
has been assumed (in accordance with the final lower bound obtained at
LEP~\cite{mhLEPfinal}). 
The MSSM parameters have been chosen in this example according to the
reference point SPS1b~\cite{sps}.
In \reffi{fig:SWMW} the inner 
(blue) areas correspond to $\de\mtexp = 0.1 \gev$ (LC), while
the outer (green) areas arise from $\de\mtexp = 2 \gev$ (LHC). 
For the error of $\De\al_{\rm had}$ we have again assumed a future
determination of $5 \times 10^{-5}$. In the SM, this is the only
relevant uncertainty apart from $\de\mt$ (the remaining effects of
future intrinsic uncertainties have been neglected in this figure). The 
future experimental uncertainty of $\mh$ is insignificant for
electroweak precision tests.
For the experimental errors on the SUSY parameters
we have assumed a 5\% uncertainty for $\mste, \mstz, \msbe, \msbz$
around their values given by SPS1b. The mixing angles in the
$\Stop$~and $\Sbot$~sectors have been left unconstrained. 
The mass of the $\cp$-odd Higgs boson $\MA$ is
assumed to be determined to about 10\%, and it is assumed that
$\tb \approx 30 \pm 4.5$, where  $\tb$ is 
the ratio of the vacuum expectation values of the two Higgs doublets of
the MSSM.

The figure shows that the improvement in $\de\mt$ from 
$\de\mt = 2 \gev$ to $\de\mt = 0.1 \gev$ strongly reduces the
parametric uncertainty in the prediction for the EWPO.  
In the SM case it leads to a reduction by about a factor of 10 
in the allowed parameter space of the $\MW-\sweff$ plane.  
In the MSSM case, where many additional parametric uncertainties enter, 
a reduction by a factor of more than 2 is obtained in this example. 
This precision will be crucial to
establish effects of new physics via EWPO.


\subsection{Indirect determination of the SM top Yukawa coupling}
\label{subsec:topyuk}

A high precision on $\mt$ is also important to obtain indirect
constraints on the top Yukawa coupling $y_t$ from EWPO~\cite{ytdet}.
The top Yukawa coupling enters the SM 
prediction of EWPO starting at \order{\al\al_t}~\cite{delrhoSMal2}.
Indirect bounds on this coupling can be obtained if one assumes that
the usual relation between the Yukawa coupling and the top quark mass,
$y_t = \wz\,\mt/v$ (where $v$ is the vacuum expectation value), is
modified. 

Assuming a precision of $\de\mt = 2 \gev$, an indirect determination
of $y_t$ with an accuracy of only about 80\% can be obtained from the EWPO
measured at an LC with GigaZ option. A precision of $\de\mt = 0.1 \gev$, 
on the other hand, leads to an accuracy of the indirect determination of 
$y_t$ of about 40\%. The inclusion of
further subleading terms beyond \order{\al\al_t} would increase this
precision; see the discussion in \cite{ytdet}. 
The indirect determination of $y_t$ from EWPO is competitive with the
indirect constraints from the $t \bar t$~threshold~\cite{mtdet2}. 
These indirect determinations of $y_t$ represent an independent 
and complementary approach to the direct measurement of $y_t$ via 
$t \bar t H$ production at the LC, which of course provides the
highest accuracy~\cite{tesla,Juste:1999af}.


\section{Implications for the MSSM}
\label{sec:MSSMimp}

\subsection{Radiative corrections in the MSSM Higgs boson sector}
\label{subsec:MSSMhiggs}

In contrast to the SM, where the Higgs boson mass is a free input parameter,
the mass of the lightest $\cp$-even Higgs boson in the MSSM can be 
predicted in terms of other parameters of the model. Thus, precision
measurements in the Higgs sector of the MSSM have the potential to play a 
similar role as the ``conventional'' EWPO for constraining the parameter
space of the model and possible effects of new physics.

At the tree level, the masses of the neutral $\cp$-even Higgs bosons
can be expressed in terms of $\MZ$, $\MA$ and $\tb = v_2/v_1$
as follows:
\BEA
m_{h, {\rm tree}}^2 &=& \edz \KKL \MA^2 + \MZ^2 
      - \sqrt{(\MA^2 + \MZ^2)^2 - 4 \MZ^2 \MA^2 \CQZb} \KKR \non \\
m_{H, {\rm tree}}^2 &=& \edz \KKL \MA^2 + \MZ^2 
      + \sqrt{(\MA^2 + \MZ^2)^2 - 4 \MZ^2 \MA^2 \CQZb} \KKR~.
\EEA
This implies an upper bound of  $m_{h, {\rm tree}} < \MZ$. 
The existence of such a bound, which does not
occur in the case of the SM Higgs boson, can be related to the fact that
the quartic term in the Higgs potential of the MSSM is given in terms of
the gauge couplings, while the quartic coupling is a free parameter in 
the SM. 

The tree-level bound, being obtained from the gauge couplings, receives
large corrections from SUSY-breaking effects in the Yukawa sector of the
theory. The leading one-loop correction is proportional to $\mt^4$. For
instance, the leading logarithmic one-loop term (for vanishing mixing
between the squarks) reads~\cite{mhiggs1l} 
\BE
\De \mh^2 = \frac{3 G_F \mt^4}{\wz\, \pi^2\,\SQb}
          \ln \KL \frac{\mste \mstz}{\mt^2} \KR~.
\EE
Corrections of this kind have dramatic effects on the predicted value of
$\mh$ and many other observables in the MSSM Higgs sector. The one-loop
corrections can shift $\mh$ by 50--100\%. Since this shift is related to
effects from a part of the theory that does not enter at tree level,
corrections even of this size do not invalidate the perturbative
treatment. In fact, the relative size of the one- and \twol\
corrections~\cite{mhiggsf1l,mhiggslong,mhiggsletter,bse,mhiggsEP2,mhiggsEP3,maulpaul,maulpaul3,mhiggsEP4,effpotfull} 
is in accordance with the expectation from a perturbative expansion.

Since these very large corrections are proportional to
the fourth power of the top quark mass, the predictions for $\mh$ and
many other observables in the MSSM Higgs sector strongly depend 
on the value of $\mt$. As a rule of thumb~\cite{tbexcl}, a shift of 
$\de \mt = 1\gev$ induces a parametric theoretical uncertainty of $\mh$
of also about $1 \gev$, i.e.\ $\De\mh^{\de\mt} \approx \de\mtexp$.

Comparing a precise measurement of $\mh$ and other Higgs sector 
observables with the MSSM prediction will allow us to obtain
sensitive constraints on the MSSM parameters, in particular $\MA$,
$\tb$, and the parameters of the stop and sbottom sectors. A precise
knowledge of $\mt$ will clearly be mandatory in order to sensitively
probe the MSSM parameters or even effects of physics beyond the MSSM. 

In order to discuss the impact of a precise measurement of $\mt$ on the 
phenomenology of the MSSM Higgs sector more quantitatively, we restrict
our analysis to the lightest MSSM Higgs boson mass $\mh$ for
simplicity. Concerning the relevance of the experimental precision of
$\mt$, as discussed above, three other sources of uncertainties have to be 
investigated, namely the experimental error, $\de\mh^{\rm exp}$, the 
parametric theoretical uncertainty from other SM input parameters, 
$\De\mh^{\rm para}$, and the intrinsic theoretical uncertainty from unknown
higher-order corrections, $\De\mh^{\rm intr}$:
\begin{itemize}

\item \ul{Experimental error:}\\
The prospective accuracies that can be obtained in the experimental
determination of $\mh$ at the LHC~\cite{mhdetLHC} and at the
LC~\cite{tesla,nlc} are:
\BEA
\de\mh^{\rm exp} &\approx& 200 \mev {\rm ~~(LHC)} \\
\de\mh^{\rm exp} &\approx& 50 \mev {\rm ~~(LC)} ~.
\EEA

\item \ul{Parametric theoretical uncertainty from other SM input
parameters:}\\
Besides $\mt$, the other SM input parameters whose experimental errors
can be relevant to the prediction of $\mh$ are $\MW$, $\als$, and
$\mb$. The $W$~boson mass $\MW$ mainly
enters via the reparameterization of the electromagnetic coupling
$\alpha$ in terms of the Fermi constant $\gf$: 
\BE
\alpha = \frac{\sqrt{2} \, \gf}{\pi}  
\MW^2 \left(1 - \frac{\MW^2}{\MZ^2}\right) 
\frac{1}{1 + \De r}~,
\label{eq:Deltar}
\EE
where the quantity $\De r$ summarizes the radiative corrections.

The present experimental error of $\de\MW^{\rm exp} = 34 \mev$ leads to
a parametric theoretical uncertainty of $\mh$ below $0.1 \gev$. In view
of the prospective improvements in the experimental accuracy of $\MW$, 
the parametric uncertainty induced by $\MW$ will be smaller than the one
induced by $\mt$, even for $\de\mt = 0.1 \gev$. 

The current experimental error of the strong coupling constant, 
$\de\als(\MZ) = 0.002$~\cite{pdg}, induces a parametric
theoretical uncertainty of $\mh$ of about $0.3 \gev$. Since a future
improvement of the error of $\als(\MZ)$ by about a factor of 2 can be
envisaged~\cite{alsdet,ewpo:gigaz1,pdg}, the parametric uncertainty
induced by $\mt$ 
will dominate over the one induced by $\als(\MZ)$ down to the level of 
$\de\mt = 0.1$--$0.2 \gev$.

The mass of the bottom quark currently has an experimental error of
about $\de\mb = 0.1 \gev$~\cite{pdg,mbhoang}. A future improvement of this
error by about a factor of 2 seems to be feasible~\cite{mbhoang}.
The influence of the bottom and sbottom  loops on $\mh$
depends on the parameter region, in particular on the values of $\tb$
and $\mu$ (the higgsino parameter). For small $\tb$ and/or $\mu$ the
contribution from bottom and 
sbottom loops to $\mh$ is typically below $1 \gev$, in which case
the uncertainty induced by the current experimental error on $\mb$ is
completely negligible. For large values of $\tb$ and $\mu$, the effect of 
bottom/sbottom loops can exceed $10 \gev$ in
$\mh$~\cite{mhiggsEP4,mhiggsAEC}. Even in these cases 
we find that the uncertainty in $\mh$ induced by $\de\mb = 0.1 \gev$
rarely exceeds the level of $0.1 \gev$, since 
higher-order QCD corrections effectively reduce the bottom quark
contributions. Thus, the parametric uncertainty induced by $\mt$ will in
general dominate over the one induced by $\mb$, even for 
$\de\mt \approx 0.1 \gev$.

\smallskip
The comparison of the parametric uncertainties of $\mh$ induced by the 
experimental errors of $\MW$, $\als(\MZ)$ and $\mb$ with the one induced
by the experimental error of the top quark mass shows that an
uncertainty of $\de\mt \approx 1 \gev$, corresponding to the
accuracy achievable at the LHC, will be the dominant parametric uncertainty
of $\mh$. The accuracy of $\de\mt \approx 0.1 \gev$ achievable at
the LC, on the other hand, will allow a reduction of the parametric
theoretical uncertainty induced by $\de\mt$ to about the same level as
the uncertainty induced by the other SM input parameters.

\item \ul{Intrinsic error:}\\
Concerning the intrinsic theoretical uncertainty in the prediction for
$\mh$ from unknown higher-order corrections, considerable progress has
been made over the last years. 
The full \onel\ corrections~\cite{mhiggsf1l}, the leading corrections at
\order{\al_t\als}~\cite{mhiggslong,mhiggsletter,bse,mhiggsEP2},
the subleading \order{\al_t^2} contributions~\cite{mhiggsEP3,maulpaul},
as well as the leading \order{\al_b\als, \al_t\al_b, \al_b^2}
corrections~\cite{mhiggsEP4} are available. 
Recently, also the full electroweak \twol\ corrections in the
approximation of vanishing external momentum have been 
published~\cite{effpotfull}.

However, at the present level of sophistication in the evaluation of
two-loop contributions to $\mh$, the intrinsic uncertainty from unknown
higher-order corrections is still estimated to be rather 
large~\cite{mhiggsAEC,feynhiggs1.2}: 
\BE
\De\mh^{\rm intr,today} \approx 3 \gev ~.
\EE
If this intrinsic uncertainty cannot drastically be reduced, it will
clearly be the dominant theoretical uncertainty in the future. On the
other hand, there are no principle obstacles that would prevent an
improvement of the accuracy of the perturbative evaluation to the 
level of $\De\mh^{\rm intr,future} \approx 0.1 \gev$. 
This very ambitious goal clearly demands an enormous effort, requiring the 
bulk part of three-loop corrections and leading higher-order contributions. 
It does not seem to be out of reach, however, on the time scale of about a
decade. 

\end{itemize}

\begin{figure}[ht!]
\begin{center}
\epsfig{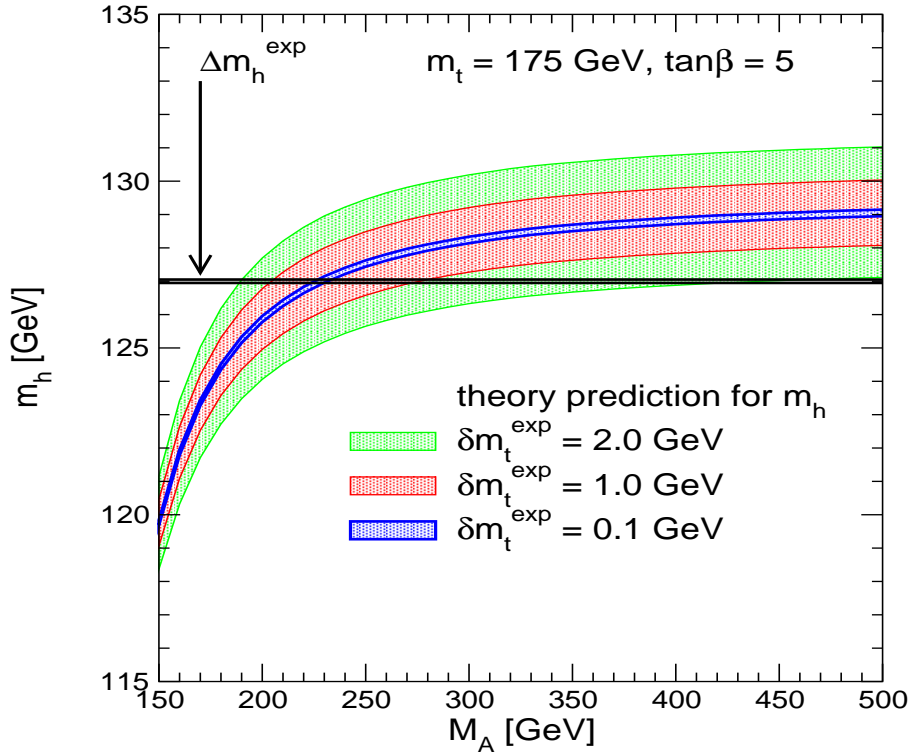}
\caption{
The prediction for $\mh$ in the $\mhmax$ scenario is shown as a
function of $\MA$ for $\mt = 175 \gev$ and $\tb = 5$. 
The three bands correspond to $\de\mtexp = 1, 2 \gev$ (LHC) and 
$\de\mtexp = 0.1 \gev$ (LC). 
The anticipated experimental error on $\mh$ at the LC
is also indicated. 
}
\label{fig:mhMA}
\end{center}
\vspace{-1em}
\end{figure}

\bigskip 
In \reffis{fig:mhMA}--\ref{fig:mhMSt2}, we focus on the experimental error 
of $\mh$ and its parametric uncertainty, while the possible impact of the 
future intrinsic uncertainty is discussed in the text.

The relevance of the parametric uncertainty in $\mh$ induced by
different experimental errors on $\mt$ is illustrated in \reffi{fig:mhMA},
where the prediction for $\mh$ is shown as a function of $\MA$ in the 
$\mhmax$ benchmark scenario~\cite{benchmark}.  
The evaluation of $\mh$ has been done with \fh~\cite{feynhiggs} for 
a central value of the top quark mass of $\mt = 175 \gev$ and for  
$\tb = 5$.
The figure shows that a reduction of the experimental error from
$\de\mtexp = 1$--$2 \gev$ (LHC) to $\de\mtexp = 0.1 \gev$ (LC) has a
drastic effect on the prediction for $\mh$.

The prospective experimental error on $\mh$ is also shown in \reffi{fig:mhMA},
while no intrinsic theoretical uncertainty from unknown
higher-order corrections is included. If this intrinsic uncertainty can
be reduced to a level of $\De\mh^{\rm intr,future} \approx 0.1 \gev$, 
its effect in the plot would be roughly as big as the one induced by 
$\de\mtexp = 0.1 \gev$. An intrinsic uncertainty of 
$\De\mh^{\rm intr,future} \approx 1 \gev$, on the other hand, would lead
to a significant widening of the band of predicted $\mh$ values (similar
to the effect of $\de\mtexp = 1 \gev$). In this case the intrinsic
uncertainty would dominate, implying that a reduction of 
$\de\mtexp = 1 \gev$ to $\de\mtexp = 0.1 \gev$ would lead to an only 
moderate improvement of the overall theoretical uncertainty of $\mh$.

Confronting the theoretical prediction for $\mh$ with a precise
measurement of the Higgs-boson mass constitutes a very sensitive test of
the MSSM, which allows us to obtain constraints on the model parameters. 
The sensitivity of the $\mh$ prediction on $\MA$ shown in \reffi{fig:mhMA}
cannot directly be translated into a prospective indirect determination
of $\MA$, however, since \reffi{fig:mhMA} shows the situation in a
particular benchmark scenario~\cite{benchmark} where, by definition,
certain fixed values of all other SUSY parameters are assumed. In a
realistic situation the anticipated experimental errors of the other
SUSY parameters, and possible effects of intrinsic theoretical uncertainties,  
have to be taken into account. In the next section, we will analyse 
the prospects for an indirect determination of SUSY parameters from 
precision physics in the MSSM Higgs sector. In particular, we will 
consider two examples of parameter determination in the stop 
sector of the MSSM.


\subsection{Constraints on the parameters of the stop sector}
\label{subsec:MSSMstop}

Once a Higgs boson compatible with the MSSM predictions has been
discovered, the dependence of $\mh$ on the top and stop sectors can
be utilized to determine unknown parameters of the
$\Stop$~sector.

The mass matrix relating the interaction eigenstates $\StopL$ and $\StopR$ 
to the mass eigenstates $\Stope$ and $\Stopz$ is given by
\BE
\label{stopmassmatrix}
{\cal M}^2_{\Stop} =
  \ML \MstL^2 + \mt^2 + \CZb (\edz - \frac{2}{3} \sw^2) \MZ^2 &
      \mt \Xt \\
      \mt \Xt &
      \MstR^2 + \mt^2 + \frac{2}{3} \CZb \sw^2 \MZ^2 
  \MR ~,
\EE
where $\Xt$ can be decomposed as $\Xt = \At - \mu/\tb$. Here, $\At$ denotes
the trilinear Higgs--$\Stop$ coupling. Assuming that $\tb$ and $\mu$ can
be determined from other sectors, there are three new parameters in the
mass matrix, $\MstL$, $\MstR$, and $\At$.
The mass eigenvalues of the stops are obtained after a rotation
by the angle $\tst$,
\BE 
{\cal M}^2_{\Stop} 
\quad
\stackrel{\tst}{\longrightarrow}
\quad
\ML \mste^2 & 0 \\ 0 & \mstz^2 \MR ~.
\EE

A possible future situation would be that the two stop quarks are
accessible at the LHC, but are too heavy for direct production at
the LC. In this case the masses $\mste$, $\mstz$ can be determined at
the LHC (supplemented with LC input~\cite{nojkawatl,lhclcdoc}), while
only limited information%
\footnote{ For low values of $\MstL$, $\MstR$ of \order{200 \gev} 
a weak sensitivity to $\tst$ might also be available from the process 
$pp \to \Stope\Stope h$ at the LHC~\cite{djouadiXt}.} 
would be obtained on the mixing angle in the stop sector, $\tst$.
Therefore the trilinear coupling $\At$ could only be loosely constrained. 
However,
the measurement of $\mh$ together with a precise determination of
$\mt$ would allow an indirect determination of $\At$. This is shown
for the benchmark scenario SPS1b~\cite{sps} in \reffi{fig:mhAt}
(evaluated with \fh). For the errors on the SUSY parameters we have
assumed a 5\% uncertainty for $\mste, \mstz, \msbe, \msbz$ around
their values given by SPS1b. $\MA$ is assumed to be determined to
about 10\%, whereas for $\tb$ a measured value of $\tb \approx 30 \pm 4.5$
is assumed. As before, we do not include intrinsic theoretical
uncertainties on $\mh$ in the plot. Its effect can most easily be
illustrated by replacing the experimental error on $\mh$ by a
combination of the experimental error and the intrinsic theoretical
uncertainty. An intrinsic theoretical uncertainty in excess of the
experimental error on $\mh$ would therefore effectively widen the
indicated interval of allowed $\mh$ values.

\Reffi{fig:mhAt} shows that the experimental error on $\mt$
has a significant impact on the indirect determination of $\At$. The
LC precision on $\mt$ gives rise to an improvement in the accuracy for
$\At$ by a factor of about 3, to be compared with the case where $\mt$
is known with the accuracy achievable at the LHC.

\begin{figure}[ht!]
\begin{center}
\epsfig{figure=mhAt02.cl.eps, width=12cm,height=10cm}
\caption{
Indirect determination of $\At$ in the SPS1b scenario for
$\de\mtexp = 2 \gev$ (LHC) and $\de\mtexp = 0.1 \gev$
(LC). 
The experimental error on the Higgs boson mass, $\de\mh^{\rm exp}$, is
indicated. For the errors on the SUSY parameters we have
assumed a 5\% uncertainty for $\mste, \mstz, \msbe, \msbz$ around
their values given by SPS1b. $\MA$ is assumed to be determined to
about 10\%, while for $\tb$ a measurement of
$\tb \approx 30 \pm 4.5$ is assumed.
}
\label{fig:mhAt}
\end{center}
\vspace{-1em}
\end{figure}

\begin{figure}[ht!]
\begin{center}
\epsfig{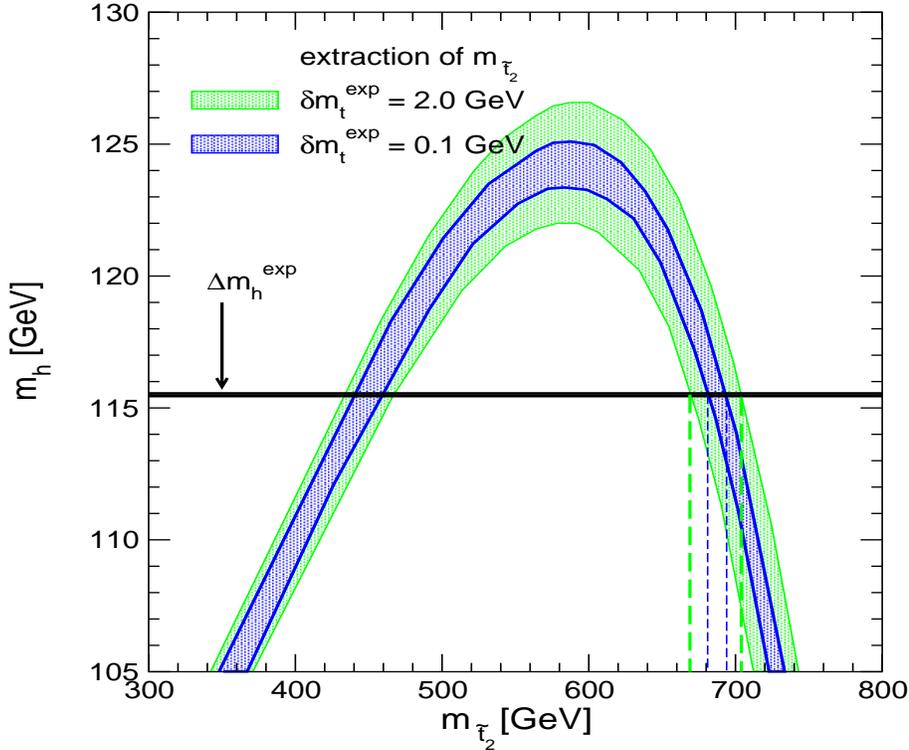}
\caption{
Indirect determination of $\mstz$ from the measurement of $\mh$ for
$\de\mtexp = 2 \gev$ (LHC) and $\de\mtexp = 0.1 \gev$ (LC). 
The experimental error on the Higgs boson mass, $\de\mh^{\rm exp}$, is
indicated. 
For the assumed experimental errors of the other SUSY parameters see text.
}
\label{fig:mhMSt2}
\end{center}
\end{figure}

As another example, in \citere{eestop} a scenario was analysed where the
lighter stop $\Stope$ is accessible at a LC with 
$\sqrt{s}=500 \gev$. 
In this case, the LC will provide precise measurements of $\mste$ and $\tst$. 
On the other hand, the heavier stop $\Stopz$ is too heavy to be 
produced at the LC in this scenario. 
In such a case, the measurement of $\mh$ can be employed for obtaining
indirect limits on $\mstz$. A comparison of this indirect determination
of $\mstz$ with a direct measurement at the LHC or a LC at higher energy 
would provide a stringent consistency test of the MSSM.

In \reffi{fig:mhMSt2} we show the allowed region in the
$\mstz$--$\mh$ plane, where the following values for the other 
parameters have been assumed~\cite{eestop,mondsichel}:
$\mste = 180 \pm 1.25 \gev$, $\costt = 0.57 \pm 0.01$, 
$\MA = 257 \pm 10 \gev$, 
$\mu = 263 \pm 1 \gev$, $\mgl = 496 \pm 10 \gev$, $\Ab = \At \pm 30\%$,
and a lower bound of $200 \gev$ has been imposed on the lighter sbottom
mass. For $\tb$ only a lower bound of $\tb > 10$ has been assumed, which
could for instance be inferred from the gaugino/higgsino sector. 
For the Higgs-boson mass a measured value of $\mh = 115.5 \pm 0.05 \gev$
has been assumed. Concerning the intrinsic theoretical uncertainty of
$\mh$, the same applies as for \reffi{fig:mhAt} above.

Intersecting the assumed measured value for $\mh$ with the allowed
region in the $\mstz$--$\mh$ plane allows an indirect determination of 
$\mstz$ in this example, yielding $670 \gev \lsim \mstz \lsim 705$ GeV
for $\de\mt = 2 \gev$ (we consider only the intersection at higher values 
of $\mstz$, since we had assumed that $\mstz$ lies above the LC reach).
This precision increases with $\de\mt = 0.1 \gev$ to $680 \gev \lsim
\mstz \lsim 695 \gev$, i.e.\ by a factor of more than 2.


\section{Renormalization group running of SUSY-breaking parameters}
\label{sec:RG}

\subsection{Neutralino and chargino masses in mSUGRA}
\label{subsec:MSSMpara}

Once SUSY particles are detected and their properties are measured,
a major task will be to determine the underlying scheme for
supersymmetry breaking out of the experimental data. For this purpose
the low-energy parameters are related to high-scale parameters
by means of renormalization group equations (RGEs).
The precision with which this can be done naturally depends on 
the amount and precision of the available experimental data. 
The top quark mass and top Yukawa coupling are important parameters 
in the renormalization group running and in radiative corrections 
to SUSY masses. 
A precise knowledge of $\mt$ will thus be very important 
for the determination of fundamental-scale SUSY parameters and  
for testing different models of SUSY breaking.  

\begin{figure}[t]
\begin{center}
\unitlength=1mm
\begin{picture}(147,72)
\put(5,2){\mbox{\epsfig{figure=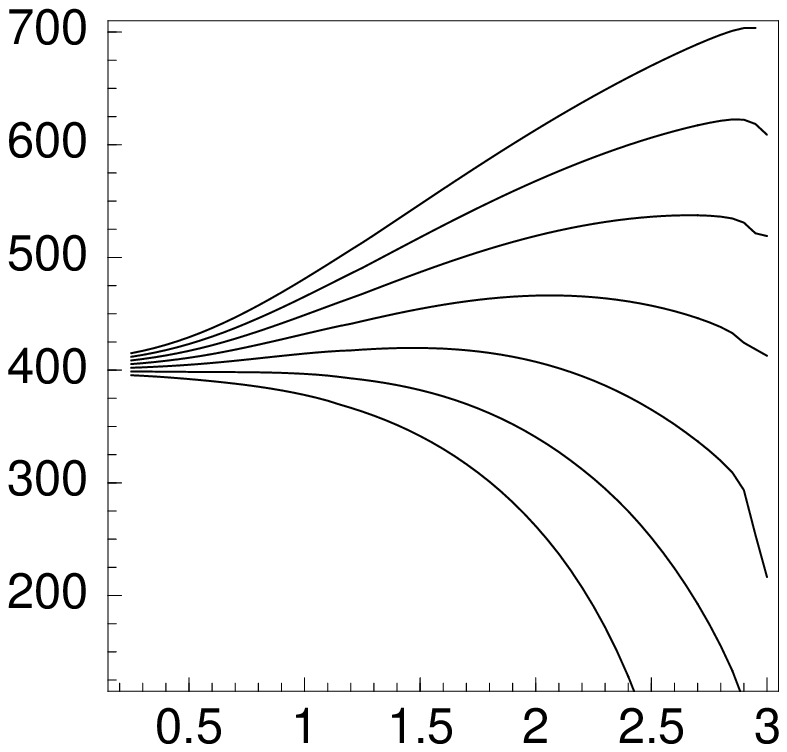,height=6.5cm}}}
\put(-4,65){\mbox{\bf (a)}}
\put( -2,30){\rotatebox{90}{$|\mu|$~[GeV]}}
\put(32,-1){\mbox{$m_0$~[TeV]}}
\put(38,59){\mbox{\footnotesize $\mt=178$}}
\put(59,57){\mbox{\footnotesize 177}}
\put(59,49){\mbox{\footnotesize 176}}
\put(60,41){\mbox{\footnotesize 175}}
\put(60,32){\mbox{\footnotesize 174}}
\put(59,22){\mbox{\footnotesize 173}}
\put(46,16){\mbox{\footnotesize 172}}
\put(79,65){\mbox{\bf (b)}}
\put(88,2){\mbox{\epsfig{figure=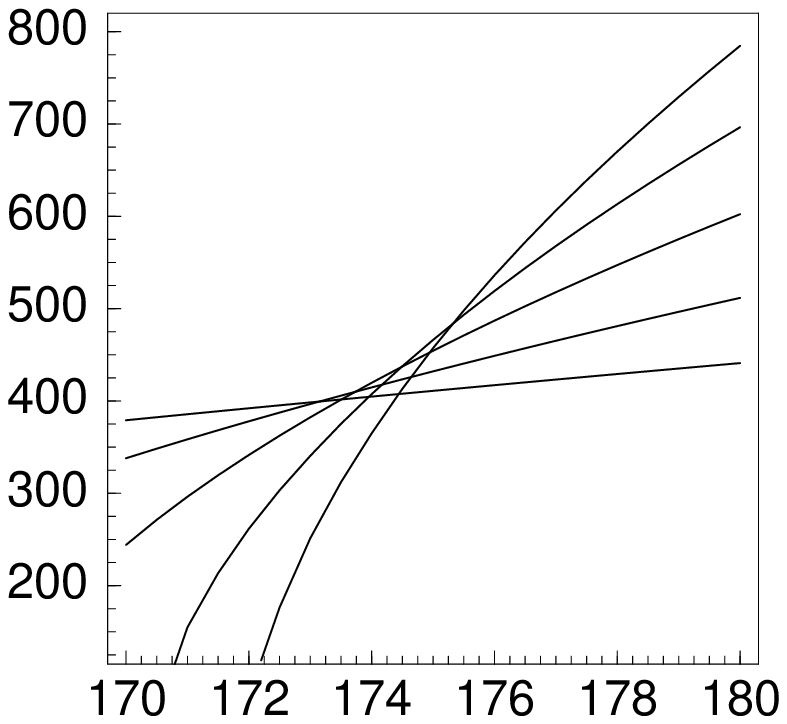,height=6.5cm}}}
\put(81,30){\rotatebox{90}{$|\mu|$~[GeV]}}
\put(115,-1){\mbox{$\mt$~[GeV]}}
\put(129,57){\mbox{\footnotesize $m_0=2.5$}}
\put(145,55){\mbox{\footnotesize 2.0}}
\put(144,47){\mbox{\footnotesize 1.5}}
\put(144,41){\mbox{\footnotesize 1.0}}
\put(144,31){\mbox{\footnotesize 0.5}}
\end{picture}
\caption{
Dependence of $\mu$ on (a) $m_0$ for various values of $\mt$ (in GeV)
and (b) $\mt$ for various values of $m_0$ (in TeV)
within the mSUGRA model, for $m_{1/2}=300 \gev$, $A_0=0$, 
$\tb=10$; calculated with SPHENO\,2.0.1~\cite{spheno}.
}
\label{fig:mumtm0}
\end{center}
\vspace{-1em}
\end{figure}

Let us consider the mSUGRA model~\cite{mSUGRA} as an illustrative example.
This model is characterized by five parameters: 
a common scalar mass $m_0$, a common fermion mass $m_{1/2}$, and 
a common trilinear coupling $A_0$ at the grand unification (GUT) scale, 
supplemented by $\tb$ and the sign of the higgsino mass parameter $\mu$.
The MSSM spectrum is determined from these five parameters by RG evolution. 
It turns out that the $\mu$ parameter is very sensitive to the top quark 
mass, especially in the case of large $m_0$.  
With a top pole mass around 174~GeV, the $\mu$ parameter  
depends only weakly on the value of $m_0$.  
Such scenarios are called focus-point scenarios \cite{focus}. In these
scenarios several large terms cancel, implying
that a small change of the top mass can lead to huge effects.
\Reffi{fig:mumtm0}a shows $\mu$ as a function of $m_0$ for 
$m_{1/2}=300 \gev$, $A_0=0$, $\tb=10$ and top quark  
masses of 172--178~GeV. 
As can be seen, the $m_0$ dependence of $\mu$ (i.e.\ the absolute value 
of $\mu$ as well as whether or not focus-point behaviour is found) 
is highly sensitive to the exact value of $\mt$. 
Analogously, \reffi{fig:mumtm0}b shows $\mu$ as a function of $\mt$ for 
several fixed values of $m_0$. Again we find a very pronounced  
dependence on $\mt$ for large $m_0$. For $m_0\sim 2 \tev$, for instance, 
a shift in $\mt$ of 1~GeV leads to a shift in $\mu$ of about 100~GeV.
In gauge- or anomaly-mediated SUSY breaking, 
the situation is somewhat more stable, as there are no 
focus-point-like scenarios. However, an error of 2~GeV on $\mt$ 
can still introduce an error of about 50~GeV for $\mu$. 
This reduces to an \order{1 \gev} error for $\de \mt = 0.1 \gev$.

The dependence of $\mu$ on $\mt$ directly translates into the 
predicted chargino and neutralino masses. This is shown in 
\Refta{tab:dmchi}, where we list the parametric errors on 
$m_{\tilde\chi^0_1}$, $m_{\tilde\chi^\pm_1}$ and 
$m_{\tilde\chi^\pm_2}$ for $\de\mt=0.1$--$2 \gev$ and two values of 
$m_0$, 0.5 and $1 \tev$.
The other parameters are $m_{1/2}=300 \gev$, $A_0=0$, 
$\tb=10$, $\mu>0$ as above, and $\mt=175 \gev$. The errors on
the chargino and neutralino
masses scale roughly linearly with the error on the top mass.

As can be seen, a precise knowledge of $\mt$ is essential for 
precise predictions. Such predictions within a particular model 
can be used for e.g., exclusion limits or consistency checks. 
For example, if the properties of $\tilde\chi^0_1$, 
$\tilde\chi^0_2$ and $\tilde\chi^\pm_1$ are measured with high 
precision, the parameters $M_1$, $M_2$ and $\mu$ can be derived  
without assuming a particular SUSY-breaking scenario~\cite{inos}. 
Combining this with information on 
the squark and gluino masses from the LHC, one may then perform a hypothesis 
test of mSUGRA (or other models) in a top--down approach, see e.g.\ 
\cite{ATLASTDR}. From \Refta{tab:dmchi} it becomes clear that 
\order{100 \mev} precision on $\mt$ is necessary for mSUGRA fits in 
order to match the experimental precision of gaugino mass measurements 
at a LC.

For completeness we note that at present there is a non-negligible 
theoretical uncertainty in SUSY mass spectrum calculations from RG running 
\cite{Allanach:2002pz}. The main source is the relation between
the measured top mass and the Yukawa coupling $y_t$. The current results
have been obtained using the complete SUSY one-loop relation plus the 
gluonic part of the two-loop relation. Recently also the SUSY-QCD part of
the two-loop relation has been given in the literature \cite{tobefound}.
In order to fully exploit the LC precision on $\mt$, further
improvements will be necessary, in particular a complete two-loop
calculation of $y_t$.

The relation between the measured top mass and the Yukawa coupling $y_t$ 
depends also on the precise value of $\als$. Shifting the value of $\als$
by $10^{-3}$ amounts to shifting $|\mu|$ by 4 GeV for $m_0=2.5 \tev$.
For large $\tb$, the error on the bottom mass has a similar influence. 
Taking  $\tb = 50$ and $m_0=2.5 \tev$ and the other parameters
as above, a shift of 0.2~GeV in the bottom mass induces a shift in 
$|\mu|$ of 5~GeV. These uncertainties are of the same order of magnitude
as those induced by the shift of $\de \mt=0.1 \gev$. For comparison, at 
$m_0=2.5 \tev$ a shift of $\de \mt=0.1 \gev$ induces a shift in $\mu$
of 8 GeV.

\begin{table}[t]
\renewcommand{\arraystretch}{1.5}
\begin{center}
\begin{tabular}{|c|rrr|rrr|}
\cline{2-7} 
 \multicolumn{1}{c|}{}   & \multicolumn{3}{c|}{$m_0=0.5 \tev$}
  & \multicolumn{3}{c|}{$m_0=1 \tev$} \\
  \cline{1-1} $\de\mt$ 
  & $\de m_{\tilde\chi^0_1}$ & $\de m_{\tilde\chi^\pm_1}$ 
                             & $\de m_{\tilde\chi^\pm_2}$
  & $\de m_{\tilde\chi^0_1}$ & $\de m_{\tilde\chi^\pm_1}$ 
                             & $\de m_{\tilde\chi^\pm_2}$ \\
\hline\hline
  0.1 & 0.01~ & 0.06            & 0.54             & 0.02 & 0.15 & 1.6 \\
  0.2 & 0.02~ & 0.11            & 1.1\hphantom{0}  & 0.05 & 0.30 & 3.1 \\
  0.5 & 0.05~ & 0.29            & 2.7\hphantom{0}  & 0.13 & 0.78 & 7.8 \\
  1.0 & 0.10~ & 0.59            & 5.5\hphantom{0}  & 0.26 & 1.7\hphantom{0}
  & 16.0 \\
  2.0 & 0.21~ & 1.2\hphantom{0} & 11.0\hphantom{0} & 0.59 & 3.7\hphantom{0}
  & 32.0 \\
\hline\hline
\end{tabular}
\end{center}
\caption{Uncertainties in predicted neutralino and chargino masses in mSUGRA 
(in GeV) due to $\de\mt$ for $m_{1/2}=300 \gev$, $A_0=0$, $\tb=10$, 
$\mu>0$ and $\mt=175 \gev$; calculated with SPheno2.0.1~\cite{spheno}. 
The central values for the masses are 
$m_{\tilde\chi^0_1}=121.02 \gev$, $m_{\tilde\chi^\pm_1}=226.53 \gev$, 
$m_{\tilde\chi^\pm_2}=435.17 \gev$ for $m_0=500 \gev$ and 
$m_{\tilde\chi^0_1}=123.06 \gev$, $m_{\tilde\chi^\pm_1}=230.97 \gev$,  
$m_{\tilde\chi^\pm_2}=456.13 \gev$ for $m_0=1 \tev$.  
For the other neutralinos, 
$m_{\tilde\chi^0_2}\sim m_{\tilde\chi^\pm_1}$,  
$m_{\tilde\chi^0_{3,4}}\sim m_{\tilde\chi^\pm_2}$
and 
$\de m_{\tilde\chi^0_2}\sim \de m_{\tilde\chi^\pm_1}$,  
$\de m_{\tilde\chi^0_{3,4}}\sim \de m_{\tilde\chi^\pm_2}$.
\label{tab:dmchi} }
\renewcommand{\arraystretch}{1}
\end{table}



\subsection{Dark matter constraints in the mSUGRA scenario}
\label{sec:dark}

\begin{figure}[htb!]
\begin{center}
\epsfig{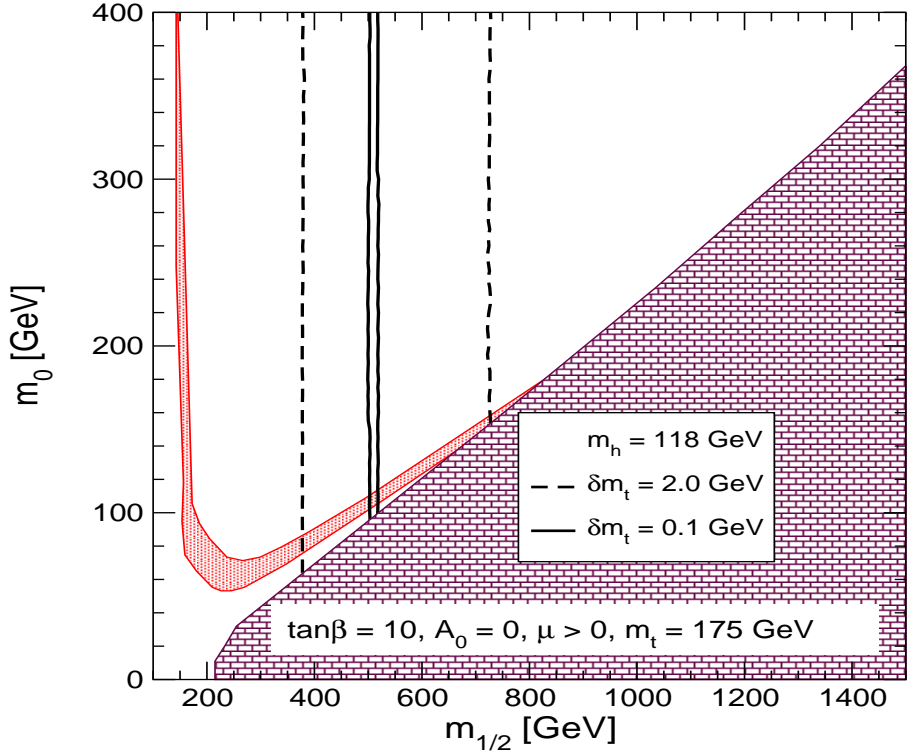}
\caption{
The CDM-allowed region from the WMAP measurement is shown in the
mSUGRA scenario in the $m_{1/2}$--$m_0$ plane for $A_0 = 0$, 
$\tb = 10$ and $\mu > 0$. The brown (bricked) region is excluded
because the LSP is the charged $\Staue$ in this region. An assumed
measurement of $\mh = 118 \pm 0.05 \gev$ is shown together with the 
parametric uncertainty
from $\de\mt = 0.1 \gev$ (solid line) and $\de\mt = 2.0 \gev$
(dashed).  
}
\label{fig:ehowtb10}
\end{center}
\end{figure}

Within the mSUGRA scenario, the lightest SUSY particle (LSP), which we
take to be the lightest neutralino, is a good candidate for cold dark
matter (CDM). After the recent WMAP measurements~\cite{wmap}, the required 
amount of CDM given by neutralinos is narrowed down to 
$0.094 \le \Omega_{\chi} h^2 \le 0.129$.
In \reffi{fig:ehowtb10} we show the CDM-allowed region in the
$m_{1/2}$--$m_0$ plane for $A_0 = 0$, $\tb = 10$ and 
$\mu > 0$~\cite{wmapdata,cdm}.
The evaluation of the CDM-allowed region has
been made for $\mt = 175 \gev$; however, no relevant change is expected
from variations in $\mt$ by $\pm 2 \gev$. Also for large $\tb$, the
CDM-allowed region is not significantly changed by variations in $\mt$ 
for the parameter space shown in \reffi{fig:ehowtb10}. If the Higgs boson 
mass is
measured, it can be used to further constrain the mSUGRA parameter
space. In \reffi{fig:ehowtb10} we have assumed an experimental
determination of $\mh = 118 \pm 0.05 \gev$. The parametric uncertainty induced
by $\de\mt$ is indicated. For $\de\mt = 2 \gev$ the allowed regions
are $400 \gev \lsim m_{1/2} \lsim 750 \gev$ and 
$80 \gev \lsim m_0 \lsim 160 \gev$. These uncertainties shrink to
$500 \gev \lsim m_{1/2} \lsim 520 \gev$ and 
$100 \gev \lsim m_0 \lsim 120 \gev$ with $\de\mt = 0.1 \gev$. This
corresponds to an improvement by factors of 17 and 4 for $m_{1/2}$ and
$m_0$, respectively.


\subsection{Bottom--up determination of SUSY-breaking parameters}
\label{subsec:bottomup}

If LHC measurements of the MSSM spectrum are complemented by high 
precision measurements at a prospective LC with sufficiently high 
energy, one can try to reconstruct the original theory at the high 
scale in a model-independent way; see \citere{Blair:2000gy,Blair2}. 
In \cite{Blair:2000gy,Blair2} it has been assumed that $\mt$ is known with 
$\de \mt=0.1 \gev$ accuracy.  In the following we will study the
situation if $\mt$ is known less precisely.

We shortly summarize the used procedure, as further details 
can be found in \cite{Blair:2000gy,Blair2,spheno}.
We take the masses and cross sections of a particular point in the SUSY
parameter space together with their expected experimental errors from the
LHC and a $\sqrt{s}=800 \gev$ LC. 
We assume that electrons can be polarized to 80\% and positrons to
40\% at the LC~\cite{polarization}. 
We then fit the underlying SUSY-breaking
parameters (at the electroweak symmetry breaking scale, $Q_{\rm EWSB} =
\sqrt{\mste\mstz}$) to these observables. An initial set
of parameters is obtained by inverting the tree-level formulas for masses
and cross sections. This set serves as a starting point for the fit, 
which is carried out with MINUIT \cite{James:1975dr} to obtain
the complete correlation matrix. In the fit, the complete spectrum 
is calculated at the \onel\ level using the formulas given 
in \cite{Pierce:1996zz}. 
In the case of the neutral Higgs bosons as well as of the $\mu$ parameter, 
the \twol\ corrections as given in \cite{mhiggsEP2,mhiggsEP3,Dedes:2002dy} 
are included. In addition, the cross sections for third-generation
sfermions at a LC are calculated (including the effect of initial-state
radiation \cite{drees1} and, in the case of squarks, also SUSY-QCD
corrections \cite{drees1,Eberl:1996wa}). 
The low-scale SUSY-breaking parameters and the errors on them are then run
up to the high-energy (GUT) scale.  In this way one can check the extent 
to which the original theory can be reconstructed. The new ingredient
with respect to \citeres{Blair:2000gy,Blair2} is that we include here the
effect of the uncertainty of the top mass and all trilinear
couplings. The effects of the $\mt$ error are: 
(i) the top Yukawa coupling gets changed and this affects the RGE running
of several other parameters, such as $y_b$, $A_t$, $M^2_{\tilde Q_3}$, 
$M^2_{\tilde U_3}$ and $M^2_{H_2}$ (the mass parameter of the 
second Higgs doublet); 
(ii) $\mt$ enters directly the calculation of the stop masses;
(iii) $\mt$ enters the loop-corrected relations between third-generation 
squark masses, chargino masses, neutralino masses, the gluino mass,
the Higgs masses and the underlying parameters in the Lagrangian.
For $\mh$, we take
$\De \mh = \sqrt{ (\de \mh^{\rm exp})^2 + (\De m^{\rm intr}_h)^2}$, 
with $\de \mh^{\rm exp} = 50 \mev$, and we use here a future intrinsic
uncertainty of $\De m^{\rm intr}_h = 0.5 \gev$.

The SUSY-breaking parameters that are most 
sensitive to $\mt$ are $A_t'$, $M^2_{\tilde Q_3}$, $M^2_{\tilde U_3}$ and 
the Higgs mass parameter $M^2_{H_2}$.
Note that we take $\At' \equiv \At \cdot y_t$, where $y_t$ is the top
Yukawa coupling, as input for
the fit, because it is this parameter that appears in the Lagrangian.

As an example we take the mSUGRA parameters
\BE
m_0 = 200 \gev, \;
m_{1/2} = 250 \gev, \;
A_0 = -100 \gev, \;
\tb = 10, \;
\mathrm{sign}(\mu) = (+)
\label{mSUGRAexample}
\EE
to generate the SUSY spectrum -- then `forgetting' this origin of
the masses and cross sections.
The value of $\At'$ at the GUT scale is in this example $-51 \gev$.
The same experimental errors as given in \cite{Blair2} are assumed.
This holds except for $\mh$ as discussed above, and 
we also assume more conservatively that the trilinear couplings in
the $\Sbot$~and $\Stau$~sector $\Ab$ and $\Atau$  
can be determined within 30\%. 
(See \citere{stauexp} for first attempts in this direction which
indicate that this might be achievable.)
This assumption does not influence the errors on the parameters
$A_t'$, $M^2_{\tilde Q_3}$ and $M^2_{\tilde U_3}$ and $M^2_{H_2}$
at the electroweak scale, but has some impact on their RG evolution
in the case of $\de \mt=0.1 \gev$:
doubling the error on $\Ab$ roughly increases the errors on 
their corresponding GUT values by about 30\%.
The impact of $\Ab$ is less pronounced for larger $\de \mt$ 
because the error on the top mass then dominates.

\Refta{tab:GUTpar} lists the central values and the 1$\sigma$ errors 
of the parameters
$A_t'$, $M^2_{\tilde Q_3}$, $M^2_{\tilde U_3}$ and $M^2_{H_2}$ 
at the electroweak scale and the GUT scale.
The errors both at $Q_{\rm EWSB}$ and at $M_{\rm GUT}$ clearly depend 
on $\de\mt$. The other parameters  are less affected by $\mt$.
The errors shown in Tab.~\ref{tab:GUTpar} are to some extent correlated,
because the stop mass parameters are not only constrained by the stop masses
and the stop cross sections, but also by $\mh$ through the higher order
corrections (see also  \refse{subsec:MSSMstop}). The asymmetry of
the errors of the scalar mass parameters squared at the 
GUT scale is due to the fact that $(A_t')^2$ enters the corresponding RGEs.

\begin{table}[th]
\renewcommand{\arraystretch}{1.5}
\begin{center}
\begin{tabular}{|c||c|c||c|c||c|c|} \cline{2-7}
 \multicolumn{1}{l||}{} & \multicolumn{2}{c||}{$\de \mt=0.1 \gev$}
& \multicolumn{2}{c||}{$\de \mt=1 \gev$} 
& \multicolumn{2}{c|}{$\de \mt=2 \gev$} \\
 \multicolumn{1}{l||}{} & $Q_{\rm EWSB}$ & $M_{\rm GUT}$ & 
                          $Q_{\rm EWSB}$ & $M_{\rm GUT}$ &
                          $Q_{\rm EWSB}$ & $M_{\rm GUT}$  \\ \hline\hline
$\At'$ & $-448 \pm 17$ & $-56 \pm 25$ 
       & $-445 \pm 23$ & $-57 \pm 38$
       & $-445 \pm 34$ & $-56 \pm 54$ \\ \hline
$M^2_{\tilde U_3}$ & $186 \pm 10$  & $44^{+30}_{-26}$
            & $189 \pm 12$  & $52^{+47}_{-30}$ 
            & $189 \pm 15$  & $54^{+64}_{-46}$ \\ 
$M^2_{\tilde Q_3}$ & $267 \pm 7$  & $42^{+20}_{-16}$ 
            & $268 \pm 8$  & $45^{+25}_{-16}$ 
            & $268 \pm 12$  & $45^{+30}_{-20}$ \\ \hline
$M^2_{H_2}$ & $-127.5 \pm 0.3$   & $42^{+34}_{-20}$ 
            & $-127.7 \pm 0.3$   & $53^{+38}_{-27}$
            & $-127.5 \pm 0.4$   & $53^{+51}_{-39}$ \\ \hline\hline
\end{tabular}
\end{center}
\caption{Absolute 
 errors on the parameters $\At'$ (in [GeV]), $M^2_{\tilde U_3}$, 
  $M^2_{\tilde Q_3}$,
  and $M^2_{H_2}$ at $Q_{\rm EWSB}$ and $M_{\rm GUT}$ (in [$10^3 \gev^2$])
 for different values of $\de \mt$. The mSUGRA point probed is defined by
 \refeq{mSUGRAexample}. The `true' values of the parameters at the GUT
  scale are $\At' = 51 \gev$,
 $M^2_{\tilde U_3} = M^2_{\tilde Q_3} = M^2_{H_2} = 40 \times 10^3 \gev^2$.}
\label{tab:GUTpar}
\renewcommand{\arraystretch}{1}
\end{table}

\begin{figure}[ht!]
\vspace{2cm}
\setlength{\unitlength}{1mm}
\begin{center}
\vspace{-2cm}
\begin{picture}(160,174)
\put(-10,-11){\mbox{\epsfig{figure=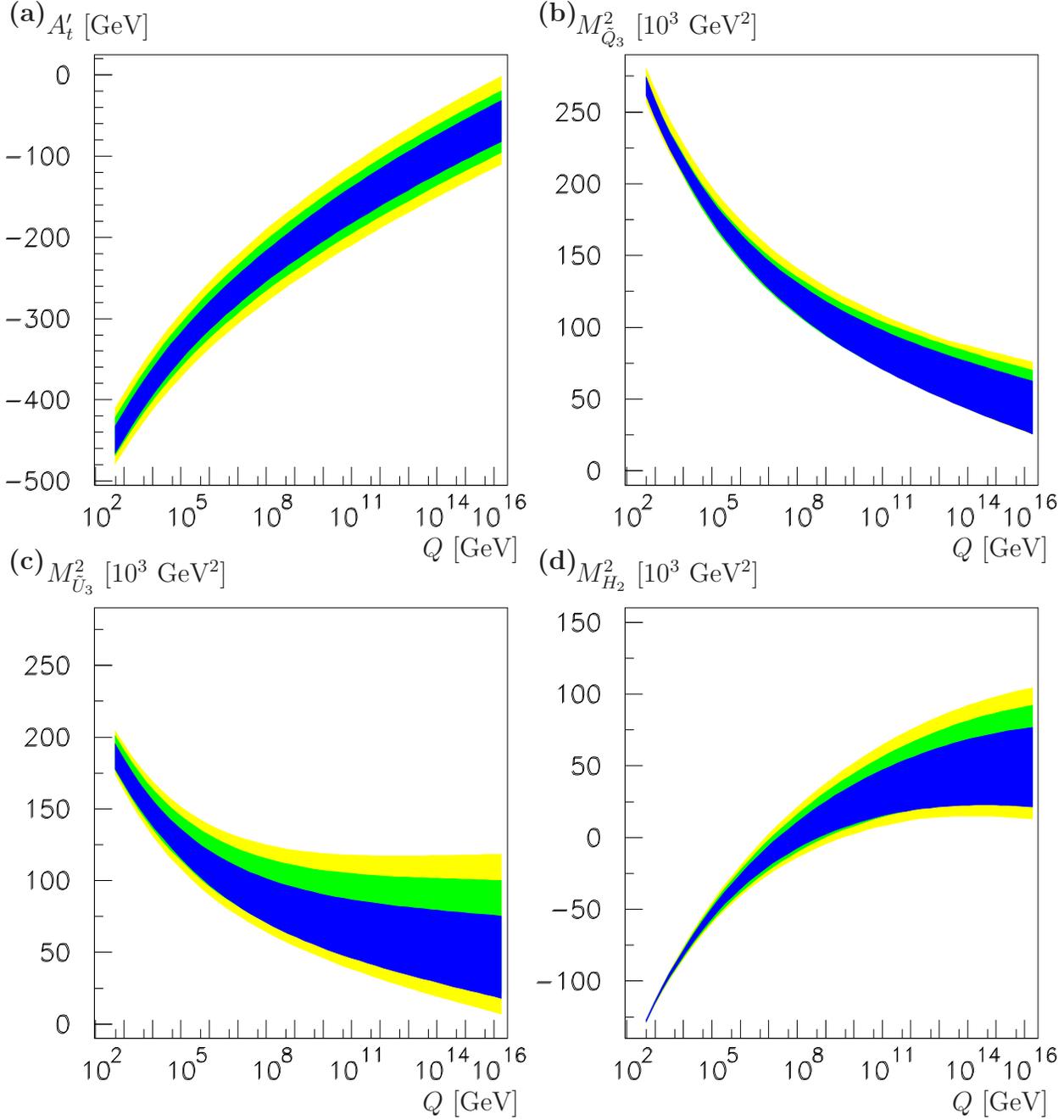,height=19.3cm,width=18.5cm}}}
\put(-5,168){\mbox{\bf (a)}}
\put(1,166){\mbox{$A'_t$~[GeV]}}
\put(60,84){\mbox{$Q$~[GeV]}}
\put(78,168){\mbox{\bf (b)}}
\put(84,166){\mbox{$M^2_{\tilde Q_3}$~[$10^3$ GeV$^2$]}}
\put(143,84){\mbox{$Q$~[GeV]}}
\put(-5,82){\mbox{\bf (c)}}
\put(1,80){\mbox{$M^2_{\tilde U_3}$~[$10^3$ GeV$^2$]}}
\put(60,-3){\mbox{$Q$~[GeV]}}
\put(78,82){\mbox{\bf (d)}}
\put(84,80){\mbox{$M^2_{H_2}$~[$10^3$ GeV$^2$]}}
\put(143,-3){\mbox{$Q$~[GeV]}}
\end{picture}
\end{center}
%
\caption{
Comparison of RGE running of (a)~$\At'$, (b)~$M^2_{\tilde Q_3}$,
(c)~$M^2_{\tilde U_3}$ and (d)~$M^2_{H_2}$, 
for the three top quark mass errors $\de\mt = 0.1 \gev$ 
(blue/dark shaded), $\de\mt = 1 \gev$ (green/medium)  and 
$\de\mt = 2 \gev$ (yellow/light).
The mSUGRA scenario is defined by \refeq{mSUGRAexample}.
The `true' values of the parameters at the GUT
scale are $\At' = 51 \gev$,
$M^2_{U_3} = M^2_{Q_3} = M^2_{H_2} = 40 \times 10^3 \gev^2$.
The widths of the bands indicate the 1$\si$ CL.
}
\label{fig:fit}
\vspace{1.5cm}
\end{figure}

In \reffi{fig:fit} we show the evolution of the parameters discussed above
for the three top quark mass errors, $\de\mt = 0.1 \gev$ 
(blue/dark shaded), $\de\mt = 1 \gev$ (green/medium)  and 
$\de\mt = 2 \gev$ (yellow/light).
It is apparent that the error on the top mass affects all four parameters.
The `asymmetric' increase of the errors is due to the different
central values of the parameters obtained in the different fits (see also 
\Refta{tab:GUTpar}) and is also a consequence of the correlations between
the different errors at the electroweak scale.

One clearly sees that $\de\mt$ should be about 0.1 GeV in
order to yield the $\Stop$~sector parameters at the high scale with a 
reasonable precision. The errors roughly increase by 60\% (100\%) if the
error of $\mt$ is increased from 0.1~GeV to 1~GeV (2~GeV). 
Moreover, the smaller the error of $\mt$, the closer the central
values of $M^2_{\tilde U_3}$, $M^2_{\tilde Q_3}$ and $M^2_{H_2}$ 
to their `true' values.
This is in particular important in view of models where 
unification of the scalar masses is imposed at the Planck scale
instead of the GUT scale: these models predict that at $M_{\rm GUT}$ 
the third-generation mass parameters as well as the Higgs mass 
parameters are different from sfermion mass parameters of the 
first two generations \cite{Polonsky:1994sr}. 

Let us finally comment on the 
influence of the errors on $\als$ and $m_b$. Shifting $\als$ by
$10^{-3}$ leads to a per-mille error on the high scale parameters.
When $\tb$ is small, an error of $0.2 \gev$ on $m_b$ induces
an error on the discussed high scale parameters, which is well below 
a per mille. For large $\tb$, the situation is more complicated because 
then also the errors on $A_b$, $M^2_{\tilde D_3}$ (the soft SUSY-breaking
parameter in the $\Sbot$~sector) and $M^2_{H_1}$ (the Higgs mass
parameter of the first doublet) play a role and the errors on the 
stop mass parameters at the high scale can easily be increased 
by several per cent. Concerning the intrinsic theoretical uncertainties
from unknown higher-order corrections, the situation is as discussed in 
\refse{subsec:MSSMpara}.


\section{Conclusions}
\label{sec:conclusions}

In this paper we have analysed the physics impact of the precise
experimental determination of the top quark mass at a prospective
$e^+e^-$~Linear Collider down to $\de\mt \lsim 100 \mev$ with respect 
to the envisaged LHC precision of $\de\mt \approx 1$--$2 \gev$.

Within the SM and the MSSM, a precise knowledge of the top quark mass
has a strong impact on the prediction of electroweak precision
observables such as $\MW$ and $\sweff$, which receive higher-order
corrections $\sim \mt^2$. 
Stringent internal consistency checks of both models are only possible
with the LC accuracy on $\mt$. In particular, a precision of $\mt$
significantly better than 1~GeV will be necessary in order to exploit
the prospective precision of the EWPO. The precise value of $\mt$
furthermore improves the indirect determination of the top Yukawa 
coupling by a factor of about 2.

The precision of the top quark mass is particularly important for
the MSSM Higgs
sector, since the parametric error of $\mh$, being $\sim \mt^4$, is
roughly given by $\De\mh^{\rm para} \approx \de\mt$. We have
demonstrated that precision physics in the MSSM Higgs sector will
require a very precise knowledge of $\mt$. The accuracy of $\de\mt =
0.1 \gev$ can, however, only be fully exploited provided that the intrinsic 
theoretical uncertainty in $\mh$ can be reduced to a similar level. 
We have also analysed indirect constraints on the parameters of the 
stop sector, taking into account anticipated future experimental 
uncertainties. In the examples we have investigated, the LC accuracy on 
$\mt$ gives rise to an improvement by a factor of 2--3 
in the indirect determination of $A_t$ or $\mstz$.

The renormalization group running of SUSY-breaking parameters is also
very sensitive to $\mt$. Within the mSUGRA scenario the error on the
predicted neutralino and chargino masses directly scales with
$\de\mt$, leading to a factor of 10 improvement when going from hadron
collider to LC precision in $\mt$.
In focus-point scenarios the dependence on $\mt$ is even more
pronounced. When the relic-density constraint is
combined with a prospective Higgs mass measurement,
the improved precision on $\mt$ leads to a significant
reduction of the allowed mSUGRA parameter space. In the bottom--up
approach for reconstructing the high-energy theory, a precise knowledge
of $\mt$ improves the uncertainty of the high-scale parameters by 
a factor of about 2.


\section*{Acknowledgements}
We thank D.J.~Miller for suggesting this project. We are grateful to
J.~Erler and M.~Winter for helpful discussions.
We thank K.~Olive for providing data for the CDM evaluation.
This work has been supported by the European Community's Human
Potential Programme under contract HPRN-CT-2000-00149 `Physics at
Colliders'.
W.P.~is supported by the `Erwin Schr\"odinger fellowship No.~J2272' 
of the `Fonds zur F\"orderung der wissenschaftlichen Forschung' of 
Austria and partly by the Swiss `Nationalfonds'.




\end{document}